\documentclass{sig-alternate-2013}
\usepackage{booktabs}
\usepackage{graphicx}
\usepackage[hyphens]{url}

\newfont{\mycrnotice}{ptmr8t at 7pt}
\newfont{\myconfname}{ptmri8t at 7pt}
\let\crnotice\mycrnotice%
\let\confname\myconfname%

\permission{Permission to make digital or hard copies of part or all
  of this work for personal or classroom use is granted without fee
  provided that copies are not made or distributed for profit or
  commercial advantage and that copies bear this notice and the full
  citation on the first page. Copyrights for third-party components of
  this work must be honored. For all other uses, contact the
  Owner/Author(s). Copyright is held by the owner/author(s).}
\conferenceinfo{WPES'15,}{October 12, 2015, Denver, Colorado, USA.}
\copyrightetc{ACM \the\acmcopyr}
\crdata{978-1-4503-3820-2/15/10. \\ DOI:
  http://dx.doi.org/10.1145/2808138.2808140}

\begin{document}

\title{On the Privacy Practices of Just Plain Sites}
\numberofauthors{5}
\author{
\alignauthor
Amirhossein Aleyasen\\
       \affaddr{University of Illinois}\\
       \affaddr{Urbana, IL, USA}\\
       \email{aleyase2@illinois.edu}
\alignauthor
Oleksii Starov \\
       \affaddr{Stony Brook University}\\
       \affaddr{Stony Brook, NY, USA}\\
       \email{Oleksii.Starov@stonybrook.edu}
\alignauthor
Alyssa Phung Au\\
       \affaddr{University of Pittsburgh}\\
       \affaddr{School of Law}\\
       \affaddr{Pittsburgh, PA, USA}\\
       \email{APA25@pitt.edu}
\and 
\alignauthor
Allan Schiffman\\
       \affaddr{Commerce Net}\\
       \affaddr{Palo Alto, CA, USA}\\
       \email{ams@commerce.net}
\alignauthor
Jeff Shrager\\
       \affaddr{\textit{Corresponding author}}\\
       \affaddr{Commerce Net}\\
       \affaddr{Palo Alto, CA, USA}\\
       \email{jshrager@stanford.edu}
}

\maketitle
\begin{abstract}

In addition to visiting popular sites such as Facebook and Google, web
users often visit more modest sites, such as those operated by
bloggers, or by local organizations such as schools. Such sites, which
we call ``Just Plain Sites'' (JPSs), are likely to inadvertently
present greater privacy risks than highly popular sites, because they
are unable to afford privacy expertise. To assess the prevalence of
the privacy risks to which JPSs may inadvertently be exposing their
visitors, we examined privacy practices that could be observed by
analysis of JPS landing pages. We found that many JPSs collect a great
deal of information from their visitors, and share a great deal of
information about their visitors with third parties. For example, we
found that an average of 7 third party organizations are informed when
a user visits a JPS. Many JPSs additionally permit a great deal of
tracking of their visitors. For example, we found that third party
cookies are used by more than 50\% of JPSs. We also found that many
JPSs use deprecated or unsafe security practices. Our goal is not to
scold JPS operators, but to raise awareness of these facts among both
JPS operators and visitors, possibly encouraging operators to take
greater care in their implementations, and visitors to take greater
care in how, when, and what they share.

\end{abstract}

\category{K.4.1}{Computers and Society}{Public Policy Issues}[Privacy]
\category{K.6.5}{Management of Computing and Information Systems}{Security and Protection}[Unauthorized access]

\keywords{Web privacy, Just Plain Sites, Third party organizations,
  Information leakage, Facebook login}

\section{Introduction}

Whereas much attention has been paid to the risks posed by the
web-based collection of private information by large organizations
such as banks, large corporations, and the government (e.g.,
\cite{Cranor2013, YoungQuan-Haase2009,
  HumphreysGillKrishnamurthy2010}), internet citizens also commonly
visit web sites operated by small organizations such as mom-and-pop
shops, blogs, sites for community activities, and school clubs and
teams. We refer to this category of sites as ``Just Plain Sites''
(JPS's), after Lave's concept of ``Just Plain
Folks''\cite{Lave1988}. Whereas large, well-funded organizations have
the resources to operate in accordance with best privacy practices,
JPSs are more likely to unintentionally expose their visitors' private
information through inappropriate actions (e.g., collecting more
private information than required), or inactions (e.g., failing to
change default email templates to hide account passwords).\footnote{We
  will generally use the term ``operator'' to mean the owner,
  operator, developer, etc. of a site.}

In the present work we analyze JPS ``front pages'', in order to assess
the prevalence of some of the privacy risks to which the operators of
JPSs may be inadvertently exposing their visitors. These are the
practices that apply at the very first encounter with a site, usually
its home page (or, more generally, its landing page). Our ``front
page'' principle accords with a ``front of the store''
metaphor. Visitors to a physical store should expect that in the front
of the store, where they are simply browsing, they needn't be concerned with
their credit card information being stolen (at least not by the
store's proprietor), whereas once they head to the metaphorical ``back
of the store'' to make a purchase or to participate in some other
interaction with the management, they have made a conscious decision
to hand over their credit cards, etc., and have, one hopes, assessed
the safety of the situation, and decided \textit{with all appropriate
  concern and knowledge} to put themselves at whatever risk they might
feel necessary. Taking this concept back to the web: Once a visitor
clicks ``checkout'' (etc.) they are, by assumption, aware of the
privacy risks entailed by this transaction. Therefore, as we are
investigating violations and issues \textit{before} that point, we
needn't go deeply into the web site, preferring instead to examine the
privacy practices evident upon loading of the home or landing page.

Numerous studies have examined the privacy policies and practices of
``specialist'' web sites -- i.e., sites whose subject-matter is
regulated, for example, government sites, sites that collect personal
data from children under thirteen, and sites that collect health or
financial data \cite{AntonEarpReese2002, Cranor2013,
  TurowAnnenbergPublicPolicy2001}. Other studies focus on the privacy
practices of the most popular sites \cite{KrishnamurthyWills,
  MalandrinoPetta, KrishnamurthyNaryshkin}. We chose to examine
``non-specialist'' sites with relatively few visitors. Also, instead
of studying the published privacy policies appearing on those sites,
we investigated what these sites actually do (or not), not merely what
they \textit{claim} to do (or not).

The web is, of course, extremely complex, and things go on all the
time that even experienced engineers are not aware of; almost
certainly the owners, operators, and proprietors of JPSs are unaware
of most of this under-the-hood activity. Becoming aware of this may
encourage them to take steps to improve their practices, or at least
to ensure that they are doing whatever they are doing with clear
knowledge of the potential risks.

\subsection{Privacy Principles for JPSs}

Recognizing the importance of online privacy, many jurisdictions have
moved to regulate the collection, storage, processing, and transfer of
personal information \cite{Directive1995, Tokson2010,
  SoloveHartzog2014}. Although the regulations in these jurisdictions
differ in detail, they all address the topics of notice (for example,
in the form of privacy policies), visitor choice in limiting the
collection and transfer of their personal data, visitor access to their
stored data, data security, and enforcement procedures \cite{US2009,
  LandesbergLevinCurtinLev1998}. These regulations apply to both
popular web sites, such as Google and Facebook, and to the less popular
ones that we have termed JPSs. In its June 1998 Report to Congress,
the Federal Trade Commission (FTC) reiterated five Fair Information
Practice Principles (FIPPs): (1) Notice/Aware\-ness; (2) Choice/Consent;
(3) Access/Participation; (4) Integrity/Security; and (5)
Enforcement/Redress \cite{LandesbergLevinCurtinLev1998}, which were
first stated by the U.S. Department of Health, Education and Welfare
in 1973, and which have been influential in the formulations of the
1980 Organization for Economic Cooperation and Development (OECD)
Privacy Guidelines and the Asia-Pacific Economic Cooperation (APEC)
Privacy Framework of 2004 \cite{SoloveHartzog2014}.

JPS operators should be aware of at least the FIPPs, OECD Guidelines,
and APEC Framework, which we summarize in terms of five principles:
1. \textit{Minimize collection}: The risk associated with
communication and storage is proportional to the amount and type of
data collected.  Collect only the information needed for specific
purposes, all of which should be made explicit in a privacy
policy, or through other means, such as labeling at the point of
collection.  2. \textit{Protect the data you collect}: Sites that
collect visitor data obviously should have reasonable data security
practices with respect to storage, disposal, and break-ins. Indeed,
the FTC requires such practices, however, enforcement is extremely
difficult and so essentially non-existent \cite{SoloveHartzog2014}, and
is likely to be even more lax regarding JPSs, which present a huge
number of tiny targets. Moreover, even among experts there is
disagreement about what ``reasonable'' means \cite{Scott2008}; the goal
posts keep changing as criminal hackers up the ante, and even expert
engineers with the best-of-intentions make mistakes. Faced with these
complexities, some clear precautions include requiring strong
passwords, encryption when transmitting and storing private data,
avoiding the long-term retention of private data by disposing of it as
soon as it has served its purpose, and developing a strong cyber-security
plan. 3. \textit{Minimize both intentional and unintentional sharing}:
Third parties such as affiliates and service providers, are likely not
held to the same privacy standards as a given site itself. Whereas it
is less likely for JPSs to have formal affiliates, it is quite common
for them to utilize service providers and third party services. For
example, many small sites utilize Google Analytics, Facebook login,
social icons such as ``Like'' buttons, and/or advertising frames
without realizing that these may be turning over information about
their visitors to third parties, most of which employ mysterious and
complicated algorithms. This is often true \textit{even without the
  objects being clicked by the visitor} (for example, on page load).
Therefore, all uses of third party services that are not necessary for
a site's core operations should be considered suspect. 4. \textit{Post
  a privacy policy that tells visitors what actually happens}: To the
extent possible, the published privacy policies should reflect what a
site actually does, not merely what is intended or required. This is
probably more difficult for JPSs operators who may not fully
understand what their site software is doing. 5. \textit{Give visitors
  choice and access}: Visitors should be able to control their own
data to the extent possible, for example, through opt-in and opt-out
buttons \cite{MilneRohm2000}, especially when a site intends to share
data with a third party \cite{MayerMitchell}. Visitors should also be
able to check, change, or delete their private data, and their entire
profile.



\section{Experiments}

With these principles in mind, we conducted a number of experiments
with the goal of describing the ``front page'' privacy practices of
Just Plain Sites.

\subsection{Overview}

A small but critical aspect of a site's visitor privacy practices can be
determined by technical analysis of the site itself on page load,
including what sort of information is collected, or at least what sort
of information is requested from the visitor.\footnote{Many less direct
  privacy-related things can also be determined on page load, for
  example, whether the site transmits information in encrypted form,
  although we did not analyze such factors.} Therefore, we set out to
characterize the information collected by forms, pages, and policies
presented in Just Plain Sites including the various trackers and
analytics, cookies, and third party content employed by JPSs. In a
second series of experiments we dug slightly deeper to examine JPS
privacy practices related to the rapidly growing practice of ``social
login'', especially regarding permissions and password storage
practices.

We begin with some terminology that we will maintain throughout. We
then develop a rough classification of JPSs, and enter into our
central analysis, that of the use of third party services and
cookies. Next we turn to first-party information explicitly collected
via web forms, analyzing the form purpose and type of information
collected, and ask whether the use of the requested information is
sensible. Following that we examine the use of third party cookies. In
the second half of the paper we change focus to login practices, and
especially the increasingly common practice of Facebook login,
examining the types of information requested by JPSs from Facebook,
and characterizing the information requested by different types of
JPSs. We also observe several sorts of bad practices being employed by
JPSs, including using deprecated methods, and passing passwords in the
clear.

\subsection{Some Terminology}

We will usually refer to typical adult individual visitors to a web
site as ``visitors'', or sometimes ``users''. The phrase ``the site''
will generally refer to the site that given visitor is explicitly
aware that they have visited. If the site communicates with other
sites (or web services of any sort that are not under control of the
site operator), we will refer to these other sites/services as ``third
parties''. Importantly, from the point of view of the visitor to the
site, there is usually no apparent distinction between content
provided by the site itself, and that provided by third parties. Here,
in large part, lies a significant source of unintended privacy risk,
because visitors usually can not tell when they are sending data to
the site operators they intended to be communicating with, or some
other, unintended, organization. Note that this is exactly the same
situation as in criminal internet piracy -- there is an unintended
third party ``listening in'' to at least some of the conversation
(although in the present case no crime is usually taking place).

\subsection{Selecting Target Sites}

We wanted to obtain a list of U.S. sites in the ``middle tier'', not
so small as to be trivial, but also not so large as to be likely to
have resources that would enable them to easily hire privacy experts
to manage their sites. The Quantcast top million list
(\url{www.quantcast.com}; accessed on Jun 30, 2014) contains sites
used in the U.S. with more than approximately 300 monthly visits. We
removed sites with a ranking greater than 50,000, thereby excluding
those having more than approximately 30,000 monthly visits. From those
remaining we dropped ''.gov'' sites, and those with ``hidden
profiles'' (per Quantcast's terminology). This left 943,489
sites. Manual inspection of 100 sites sampled uniformly suggested that
this process resulted in a selection of sites that roughly agreed with
our sense of what a ``Just Plain Site'' should be. All of our
experiments began with this list of nearly 1 million sites.

\subsection{Rough Classification of JPSs}

Because it is likely that different sorts of sites operate under
different privacy regimens, it is useful to classify sites into
categories that are likely to accord with such regimens.  There are
various ways to categorize web sites. Existing solutions mostly
perform text/document categorization (e.g., \textbf{similarweb}), or
utilize user-provided classes (e.g., \textbf{dmoz}). We found both of
these to be too detailed for an analysis that covers such a large
number of sites. We developed a classification based upon what product
or service is being provided (content v. good), and who is providing
it (the site operator v. visitors), as described in Table
\ref{SITECATTABLE}.

\newcommand{\specialcell}[2][c]{%
  \begin{tabular}[#1]{@{}c@{}}#2\end{tabular}}

\begin{table}
\centering
\caption{Categories of JPSs and Representation}
\label{SITECATTABLE} 
\begin{tabular}{|p{1.5cm}||p{2.2cm}|p{2.2cm}|} \hline
&\specialcell{Content,\\Web Service}&\specialcell{Goods,\\Physical Service} \\ \hline\hline
\specialcell{Single\\Producer}&40\% - Blog, official site, web tool, announcements for a club, news&44\% - Online shop, poster site of a shop, paid online game \\ \hline
\specialcell{User\\produced}&3\% - Social network, forum, wiki, group discussions&Ad networks, consumer to consumer platforms \\ \hline
\hline\end{tabular}
\end{table}

We manually classified the 100 uniformly sampled sites. The
percentages in Table \ref{SITECATTABLE} represent the distribution of
classes in this set. 13\% were uninterpretable. Shopping-type sites
dominate, but there are many blog-type sites as well. Small ad
networks, like Craigslist or Ebay (but smaller) are
rare. Surprisingly, social network- and forum-type sites are
relatively uncommon.

\pagebreak
\subsection{Third Party Service Analysis}

Sites often interact with third parties for various purposes, such as
requesting static resources (e.g., code libraries, images, css,
fonts), analytics, ads, social widgets, web beacons, and so on. It is
generally difficult or impossible for the site to determine the extent
of a third party's privacy practices. When someone visits a page that
accesses third party services, some information (e.g., IP address,
browser type -- so-called ``fingerprinting'' \cite{Nikiforakis2013}))
is sent to the third party service provider, \textit{often without any
  action required by the visitor}. Usually third parties use this
information for personalized ads, improving their services, or
aggregate them with other information. Many web pages contain
``social'' buttons to ``like'', ``tweet'', ``digg'', etc., and such
trackers often add third party cookies when you visit the page, again
\textit{without any action required by the visitor}. Moreover, some of
these third parties are better ``ninjas'' than others, not leaving
tracks such as cookies, but nonetheless gaining access to the
visitor's information. For example, Facebook states that: ``If you've
previously received a cookie from Facebook because you either have an
account or have visited \url{facebook.com}, your browser sends us
information about this cookie \textit{when you visit a site with the
  ``Like'' button or another social plugin}.'' (emphasis added) Note
that the specific way that third parties use cookies will vary.  The
Facebook Like button, for example does \textit{not} install a cookie
on load, it just checks for an existing cookie from previous visits to
the Facebook site itself.  The Twitter Tweet button, by contrast,
installs a cookie on the visitor's browser.

We define a ``tracker'' as any process that the site, or a third
party, may use that would create an \textit{unintentional} privacy
risk, for example, identifying the visitor directly via a browser
profile.\footnote{This definition excludes primary usage tracking, as
  that would be intentional.} However, many commonly employed
mechanisms pose an unintentional privacy risk. For example, unless the
site operators goes out of their way to change the default behavior of
the Apache server, it will create unencrypted log files of all access
to the site, left in a commonly known default location. Static
``Side-loading'' of files (scripts, images, etc.) from third parties
is a very common practice and creates an unintentional privacy risk by
virtue of the visitor unknowingly accessing these third party sites,
thereby permitting the third party to profile the visitor who is
unwittingly accessing their site for this side-loading. Another very
common example of this sort of tracking is the use of analytic sites,
such as Google Analytics. Google, of course, utilizes this data for
their own unknowable purposes, as well as for the intended purpose of
the originating site. More insidious are ``beacons'', pieces of code
that activate when the web page is loaded, \textit{regardless of the
  visitor clicking on them}.  These are often essentially invisible --
tiny black images or space characters (so-called ``pixel tags'') --
and may be linked to arbitrary URLs or complex javascript code,
enabling data-gathering computations without visitor awareness.

\subsection{Data Capture and Cleaning}

As mentioned in the introduction, we are concerned primarily with
privacy practices that affect a visitor by just loading the site's
landing page, and by clicking the available buttons. In accordance
with this policy, we do not analyze referred domains unless there is
an immediate redirect, and then we only follow one such
redirect. Also, we only consider what is collected while the visitor
is browsing, that is, we do not fill in info required to take whatever
``next steps'' might be possible from a page. In fact, we never fill
in any information at all, but merely look at what happens upon
clicking the available buttons.

We used PhantomJS to capture all http requests and responses on page
load, ignoring ``local'' requests/responses (i.e., those within the
same domain or a subdomain). We parsed the URLs into the domain, path,
filename, and other information, and then combined URLs that appeared
to be served by the same entity (for example, \url{s1.criteo.com} and
\url{s2.criteo.com}, or where the IP addresses are the same except for
the last octet).

\subsection{General Description of the Dataset}

\begin{figure}[ht!]
\centering
\includegraphics[width=90mm]{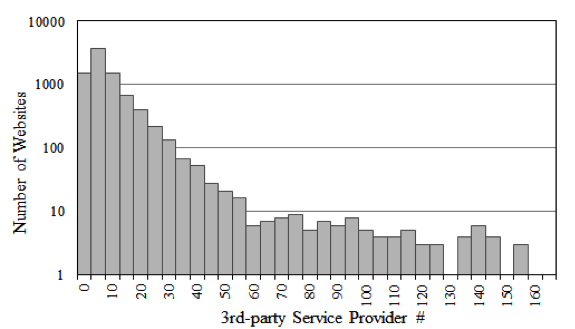}
\caption{Third party accesses per service provider}
\label{THIRDPARTYHISTORGRAM}
\end{figure}

Table \ref{THIRDPARTYSTATSTABLE} and Figure \ref{THIRDPARTYHISTORGRAM}
provide a general sense of our data. Among the 8,601 URLs accessed,
8,451 were analyzable\footnote{The remainder failed for various
  complex, uninteresting, and/or inexplicable reasons.}. Among these,
82\% had at least one third party access on page load. The number of
third party requests from a single JPS page load ranges from zero to
more than 150 (Figure \ref{THIRDPARTYHISTORGRAM}). \textit{When
  someone visits a site, an average of 7 other organizations may know
  that they have been there.}

\begin{table}
\centering
\caption{Third party accesses}
\begin{tabular}{|c|c|} \hline
Description&Count\\ \hline\hline
Requests&252036\\ \hline
Responses&251802\\ \hline
URLs accessed&503838\\ \hline
Distinct URLs&201628\\ \hline
Distinct hosts&9580\\ \hline
Distinct hosts (combined)&8601\\ \hline 
\hline\end{tabular}
\label{THIRDPARTYSTATSTABLE}
\end{table}

\subsection{Monopoly in Third Party Services}

By combining the results from different sources, a third party
organization can obtain significant additional information about a web
site's visitors. For example, if a third party provider can trace a
unique visitor across different sites by aggregating the data it is
given ``on the side'' (e.g., via browser profiles), it can better
estimate a visitor's interests and can improve targeted
advertising. Therefore, we sought to identify the owner of third party
services in our dataset.

To accomplish this we categorized URLs based on their hostname, after
truncating subdomains. Since third party service providers often use
several different domains, we had to aggregate the domains by each
owner. This task is not straightforward, and we could not develop a
fully automatic method. Therefore we manually aggregated the domains,
considering issues such as mergers and acquisitions, and
``de-referencing'' Content Delivery Networks (CDNs). For example,
Google uses numerous fronts, such as \url{doubleclick.net},
\url{gstatic.com}, \url{ytimg.com}, and \url{blogger.com}. And
\url{fbcdn.net}, as well as any subdomain of \url{akamaihd.net} that
contains ``fbcdn'' are fronts for Facebook.

Unsurprisingly, more than 67\% of the JPS web sites use at least one
service of Google. Another 19\% use at least one service of Facebook,
and about 11\% use services from Twitter. Approximately 35\% is
divided among various smaller players, like Amazon, Quantcast, and
Wordpress at about 4\%. (This may sum to more than 100\% because 
sites may use third party services from multiple providers.)

\subsection{The Purposes of Third Party Requests}

Some kinds of third party requests present more of a privacy risky
than others. For example, side-loading ``show\_ads.js'', being a
script, is almost certainly more dangerous than a style sheet or image
fetch \cite{NikiforakisInvernizzi}. We sought to characterize in the
purpose of third party requests to the top two players: Google and
Facebook. We considered combinations of several criteria such as
domain (and subdomains), path/filename, and URL parameters. It is
usually not possible to correctly recognize the purpose of a third
party request just from the domain or filename. For example, Google
Analytics uses ''\_\_utm.gif''. The content of this tiny image is
merely a ``beacon'', a great deal of information, often comprising
hundreds of bytes, is passed in the request URL.

We first recognized static resources and then worked on ads,
analytics, and other third party requests, as their recognition is
more challenging. Static resource URLs usually have no parameters, or
have only short, simple parameters, and usually do not set cookies in
their headers. The static categories that we found included static CSS
(.css), images (.jpeg, .jpg, .png, .gif), javascript (.js), json,
(.json), and html (.html).

In addition to the above static URLs, we observed these (apparent)
functions\footnote{As only the author really knows what the javascript really
  does, we characterize only what it \textit{appears} to do.}: \textit{Ads}: if
the domain is for an advertising company and there are related
keywords in the subdomain (e.g. \url{ads.yahoo.com}) or in the path
(e.g. \url{doubleclick.net/pagead/ads}); \textit{Analytics}: if the
domain is for an analytics company and there are related keywords in
the subdomain (e.g. \url{analytics.bigcommerce.com}) or in the path
(e.g. \url{.../track}); \textit{Beacons}: if there are keywords in the
subdomain (e.g. \url{pixel.quantserve.com}) or in the path
(e.g. \url{.../bug/pic.gif}) and the URL parameters are not empty; and
\textit{Widgets}: if there are keywords in the subdomain
(e.g. \url{widgets.wp.com}) or in the path
(e.g. \url{.../js/plusone.js}).

\begin{figure}[ht!]
\centering
\includegraphics[width=90mm]{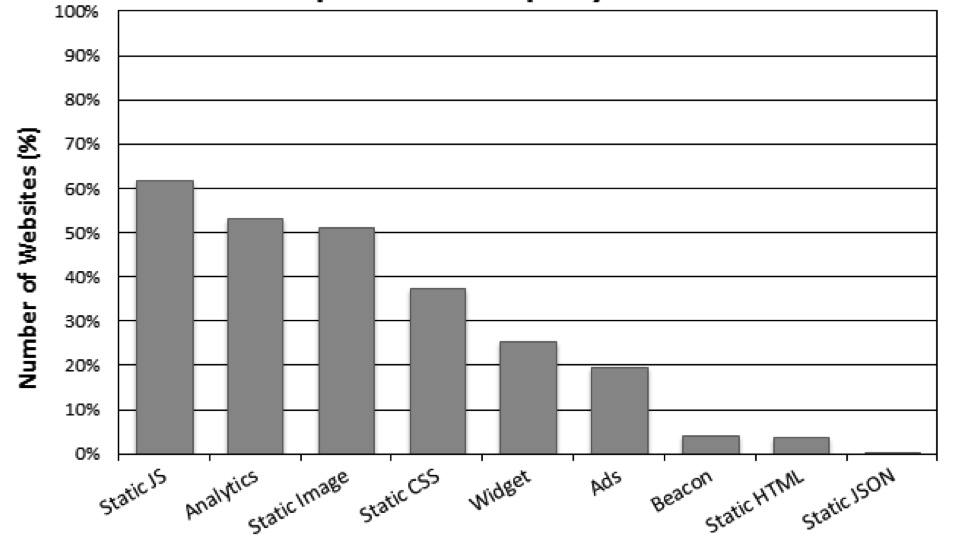}
\caption{Purpose of third party requests}
\label{PURPOSEOFTHIRDPARTYHISTO}
\end{figure}

Figure \ref{PURPOSEOFTHIRDPARTYHISTO} depicts the distribution of
usage patterns of third party requests.

\begin{figure}[ht!]
\centering
\includegraphics[width=90mm]{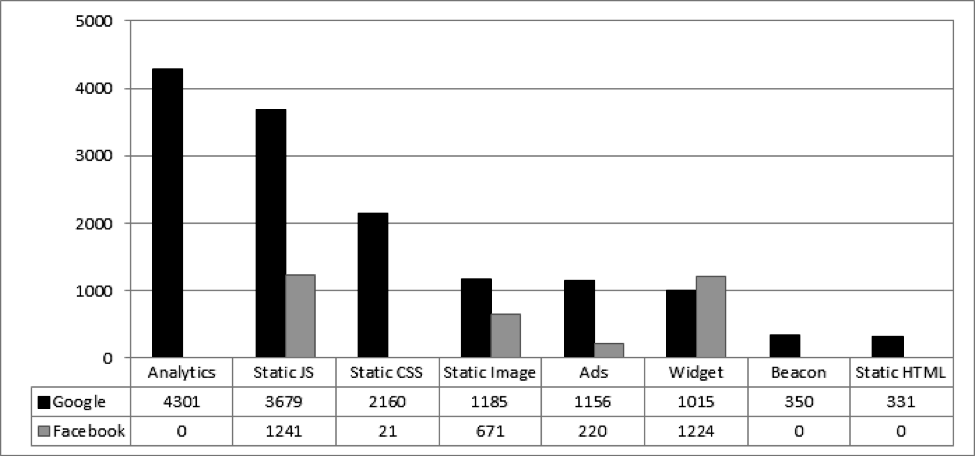}
\caption{Usage pattern of Google v. Facebook services}
\label{COMPARISONGOOGLEFB}
\end{figure}

Figure \ref{COMPARISONGOOGLEFB} depicts the usage pattern of Google
and Facebook, the two biggest third party service providers. The most
popular Google service is analytics. In terms of providing widgets, the
popularity of Google and Facebook is almost the same.

\section{First-Party Forms}

So far we have been discussing only implicitly-collected information,
but much more invasive information, such as so-called ``Personally
Identifiable Information'' (PII), is often explicitly requested via
html forms on the landing page. In order for a visitor to decide
whether the personal information being requested is necessary, it is
necessary to know the purpose to which the information will be put.
For example, it should be a red flag for a site to request your
physical mailing address in order to subscribe to an electronic
newsletter. There \textit{may} be a rational reason for this, but it
would be useful to have an explanation before making the choice about
whether to reveal this sort of information.

\subsection{Web Form Purpose and Information Type}

\begin{figure}[ht!]
\centering
\includegraphics[width=90mm]{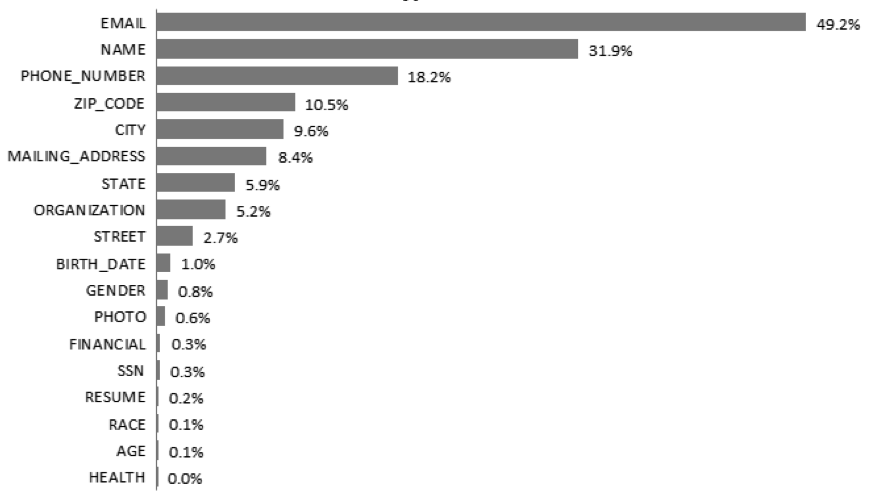}
\caption{Information collected in web forms}
\label{WebFormInformationTypeSIDEHIST}
\end{figure}

According to our analysis, 54\% of JPSs collect visitors' PII through
HTML forms. We tried to classify the \textbf{purpose} of web forms and
the \textbf{information type} of their fields.  In order to classify
the information type of the form (Figure
\ref{WebFormInformationTypeSIDEHIST}) we considered the name, label,
and default value of the fields. When a field had no label we
considered the text of a previous sibling if a label was available
there. Finally, if we still could not classify the information
requested, we checked the first parent of the DOM element and also the
text in the field, if any. In order to classify the purpose of the
form (Figure \label{PURPOSEOFWEBFORMSIDEHIST}), we considered the text
of the first child of the form (hopefully a title or short description
of the form's purpose), the previous sibling of the form (often a
title for the form is there), and also the text body surrounding the
form, if it was not unreasonably long. Table \ref{COLORFULTABLE}
depicts the type-by-form results.


\begin{figure}[ht!]
\centering
\includegraphics[width=90mm]{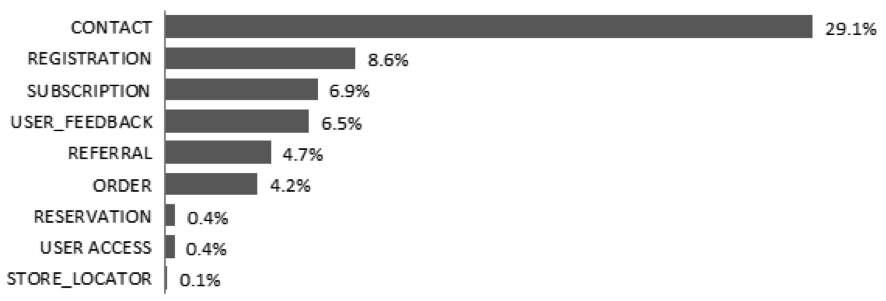}
\caption{Purpose of collected information}
\end{figure}

\begin{table*}[p!]
\centering
\caption{Distribution of PII types based on form type}
\begin{tabular}{|c||c|c|c|c|c|c|c|c|c|c|} \hline
&EMail&Name&Phone&Zip&City&Addr.&St.&Org.&St.&Bd. \\ \hline\hline
Contact&96\%&53\%&31\%&17\%&16\%&16\%&11\%&8\%&5\%&1\% \\ \hline
Order&83\%&43\%&39\%&19\%&14\%&9\%&7\%&14\%&5\%&3\% \\ \hline
Referral&84\%&50\%&12\%&13\%&7\%&4\%&3\%&4\%&2\%&1\% \\ \hline
Registration&43\%&12\%&6\%&3\%&3\%&2\%&1\%&2\%&0\%&0\% \\ \hline
Reservation&79\%&45\%&55\%&18\%&13\%&13\%&8\%&3\%&0\%&0\% \\ \hline
Store Locator&11\%&0\%&11\%&56\%&22\%&22\%&11\%&0\%&11\%&0\% \\ \hline
Subscription&95\%&31\%&7\%&6\%&4\%&4\%&2\%&2\%&1\%&0\% \\ \hline
Unknown&79\%&48\%&26\%&16\%&14\%&10\%&7\%&7\%&4\%&2\% \\ \hline
User Access&34\%&17\%&11\%&14\%&9\%&6\%&6\%&6\%&6\%&3\% \\ \hline
User Feedback&94\%&69\%&32\%&4\%&4\%&3\%&2\%&5\%&1\%&0\% \\ \hline
\hline\end{tabular}
\label{COLORFULTABLE}
\end{table*}

In order to assess the quality of our method, we manually coded a
sample of 100 forms. In fifty-eight (58\%) of these, our automatic and
manual analyses matched exactly. Among the rest, there were 131
errors, both false positive and false negative, totaling approximately
6.24\%. 


\pagebreak
\subsection{Uses of Requested Information}

Unsurprisingly, visitor names and email addresses are the most
commonly requested information. These are used for many purposes such
as registration, feeds, etc. After these, the most-collected PII is
phone numbers, and a smaller number of sites collect other PII, such
as birthdates.

We wanted to sample the reasons for birthdate being collected, and to
assess whether it seemed to make sense to collect this information,
given the purpose of the site. Table \ref{EXAMPLESTABLE} indicates the
results of this exploration. In most cases where birthdate was
collected we could determine whether it was collected in full or
partially, and whether the purpose for collecting such information
seemed to make sense.  For example, it clearly makes sense for an
insurance company to want to know an applicant's birthdate. Similarly,
it makes sense for a ``men's supper club'', to want to know a partial
birthdate, which would indicate the registrant's age, although not his
or her exact birthday. These are both sensible uses of information. On
the other hand, it makes less sense for a summer camp to require full
birthdate information; it may need to know the applicant's age, but
then it could request partial birthdate information, or equivalently,
simply request the applicant's age. Table \ref{EXAMPLESTABLE} also
indicates whether the site explicitly indicated the reason for
collecting this data. Notice that \textit{in almost no case} was this
explicitly stated, although in a few we deemed it obvious.

\begin{table*}
\centering
\caption{Manually examined examples of birthdate requests}
\begin{tabular}{|p{3cm}|p{1.5cm}|p{2.5cm}|p{2cm}|p{2.5cm}|} \hline
Page Purpose&Full BD?&Purpose&Rational?&Explained? \\ \hline\hline
Life Insurance Quote&Full&Need for Process&Yes&No but Obvious\\\hline
Heath Insurance Quote&Full&Need for Process&Yes&No but Obvious\\\hline
Auto Insurance Quote&Full&Need for Process&Yes&No\\\hline
International Student Request Info&Full&Identity&Yes&No\\\hline
Loan Information Request&Full&-&No&No\\\hline
Subscription&Partial&Special Offers&Yes&No\\\hline
Special Offer Subscription&Full&Check Age&Yes&Yes\\\hline
Join Club&Full&Check Age&Yes&Yes\\\hline
Alumni Information Update&Full&DoubleCheck Identity&?&No\\\hline
Apply for Graduate Program&Full&Identity&Yes&No\\\hline
Subscription Information Card&Full&-&No&No but Obvious\\\hline
Camp Registration&Full&-&No\; Just needs age&Yes\\\hline
Special Offer Subscription&Partial&Special Offers&Yes&No\\\hline
Funeral Arrangement&Full&-&Yes&Yes\\\hline
Membership&Full&-&No&No\\\hline
Membership&&Need for Process&Yes&No\\\hline
Subscription&Full&Need for Process&Yes&No\\\hline
For Magical Spells&Full&Need for Magical Process&Yes&No\\\hline
Free Spell Consultation Form&Full&Need for Process&Yes&No\\\hline
Application Form&Full&-&No&No\\\hline
\hline\end{tabular}
\label{EXAMPLESTABLE}
\end{table*}

\newcommand{\tpc}{third party cookie}
\newcommand{\tpcs}{third party cookies}
\newcommand{\Tpcs}{Third party cookies}

\section{Third Party Cookies}

As mentioned above, a specific sort of tracking is represented by
\tpcs, which many sites leave on visitors' browsers. We conducted a
survey of \tpcs\ based upon two uniform random samples of 1000 and
10000 JPS from our original dataset. As in the previous experiments,
we used PhantomJs to capture external resources fetched during webpage
load, including Javascript libraries, images, css files, trackers, and
ad network referrals. We evaluated how many web sites use
\tpcs\ (around 50\%), the lifetime of \tpcs\ in comparison to ``own
cookies'' (those from the visited site), and the popularity of various
\tpc\ domains. Unless otherwise stated, the following results arise
from the 10k dataset.

Those domains providing more than 5\% of \tpcs\ were (rounded):
\url{.doubleclick.net}: 18\%, \url{.google.com}: 9\%,
\url{.twitter.com}: 7\%, \url{.score-card-research.com}: 7\%,
\url{.adnxs.com}: 7\%, and \url{.youtube.com}: 6\%.  One can see that
most \tpcs\ come from ad or social networks. The lifetimes of cookies
were not very variable: most cookies' lifetimes are over a year (own:
52\% v. 3rd: 33\%). Next to this, most own cookies have ``session''
durations (own: 8\% v. 3rd: 2\%), whereas most \tpcs\ have one hour
durations (own: 1\% v. 3rd: 6\%), followed by 1 year durations (own:
2\% v. 3rd: 5\%).

\section{Third Party Logins}

Third party login, such as ``signup/login with Facebook'', is
becoming increasingly popular among JPSs. Most visitors are presumably
aware that some information is shared between the local site and
Facebook when such a login is employed, at least when specific
permissions are asked in a dialog box. However, visitors are probably not
aware of the details of this sharing.\footnote{Several studies, e.g.,
  \cite{MarkusHubery}, have studied the permissions requested by
  Facebook apps, but not local web sites that use Facebook login.}

Detecting Facebook login on a site is complicated by the variety of
ways in which it can be implemented. However, it is easier to detect
automatically in comparison with, for example, OpenID with a variety
of interfaces, or Janrain with Ajax implementations, because the code
for Facebook login is highly determined by Facebook, so we can detect
it by checking for a specific URL in a dialog window, or in the page
after clicking login links or buttons on the web site.

For this analysis we uniformly selected a subset of 100,000 (100k)
sites from our original dataset. In accordance with our assumptions
about the simplicity of JPS sites, we assumed that a
login/registration page would be accessed by at most one click from
the landing page. Web Site operators wishing to permit Facebook login
may either use the Facebook-provided html, or may code their own
custom implementation. We used PhantomJSDriver and FirefoxDriver to
render the DOM and click on various controls and tags. We targeted
XPath: ``//input[ @type= 'submit'] | //a | //button'' - for custom
implementations and ``//iframe[ @title= 'fb:login\_button Facebook
  Social Plugin'] | //div[ @class= 'fb-login-button']'' - for Facebook
login buttons, and considered only tags with login-related text such
as: ``signup'', ``connect with'', ``facebook'', etc. Although this
method does not take into account complex Ajax interactions, state
information is still saved by the target URLs. We used Selenium to
click even on hidden Facebook login buttons, if they were present in
the page source. This covers situations where the DOM is changed, for
example revealing a hidden div that appears on clicking ``signup''.

We ended up with two sets of sites that use Facebook login: 1,191
sites obtained from the BuiltWith statistics, and 260 additional sites
that use a custom Facebook login implementation, found in our 100k
sample. From these two sets (1,191 \& 260) we asked (a) what specific
information is requested by Facebook, and (b) how many sites still
generate and explicitly store passwords, in comparison to more secure
means such as sending an activation link?\footnote{Social login can
  also be done via third party platforms, such as
  \url{eventbrite.com}. We did not restrict redirections for
  signup/login.}  The second question was answered by analyzing
``congratulating new user'' emails received after login, where we also
noticed the use of several problematic practices. For all of the above
we manually checked 20 random web sites from each sample.

\subsection{Information Requested from Facebook}

Detailed results are provided in Table \ref{CUSTOMVFBLOGINTABLE},
giving the top 10 specific permissions are requests.  Naturally,
specific permissions are requested more often in the case of a custom
implementation of Facebook login: only 8\% web sites use default
``public profile'' permissions, whereas 31\% of "ready to copy-paste"
Facebook login buttons use the defaults. More interesting are the
exact permissions JPSs ask for, which are in some situations,
redundant. For example, ``user\_birthday'' is very popular, despite
the Facebook's guidelines: ``Use any available public profile
information before asking for a permission. For example, there is an
age\_range field included in the public profile...'' Also, some JPSs
are still using deprecated permissions, such as ``offline\_access''.

\newcommand{\ra}[1]{\renewcommand{\arraystretch}{#1}}
\begin{table*}\centering
\caption{Information requests at Facebook login (Left over 260 sites, Right over 1,191 sites)}
\begin{tabular}{|r|r|c|r|r|}\toprule
\multicolumn{2}{c}{Custom Implementation} & \phantom{abc} & \multicolumn{2}{c}{Login Button} \\
\cmidrule{1-2} \cmidrule{4-5} 
Permission & Percentage && Permission & Percentage\\ 
\cmidrule{1-2} \cmidrule{4-5} 
email&90\%&&email&67\%\\
user\_birthday&28\%&&user\_birthday&33\%\\
publish\_stream&23\%&&<default>&31\%\\
user\_location&17\%&&publish\_stream&22\%\\
read\_stream&10\%&&(offline\_access)&13\%\\
user\_about\_me&9\%&&user\_about\_me&12\%\\
(offline\_access)&9\%&&user\_location&10\%\\
user\_likes&9\%&&user\_likes&8\%\\
<default>&8\%&&read\_stream&7\%\\
publish\_actions&8\%&&status\_update&5\%\\
\cmidrule{1-2} \cmidrule{4-5} 
\end{tabular}
\label{CUSTOMVFBLOGINTABLE}
\end{table*}

\subsection{Clustering for Facebook JPSs}

As expected, Facebook login permission requests differ for different
categories of web site, and at the same time result in natural
clusters.  According to our four-way categorization, depicted in
Figure \ref{SITECATTABLE}, among the 260 sites in the smaller sample,
48\% were shopping sites, 34\% were blogs, 17\% were social, and 1\%
were ad sites. We decided to explore organic clustering based upon the
permissions requested by these sites, as described above. We began
with the 48 sites requesting Facebook permissions from those requested
by JPSs with custom Facebook login code (from the experiment with 100k
samples). We first combined the permissions requested by less than 5\%
of the sites into a single attribute ``other''.\footnote{If the site
  requested \textit{only} default permissions (actually, asking for a
  public profile), we did not consider ``default'' as a separate
  attribute.}  We clustered the resulting data using Weka's XMean
algorithm. This resulted in a stable clustering based upon 13
features. The final clusters, ordered by amount of information
requested, are summarized in Table \ref{AMOUNTOFINFOTABLE} (next
page).

\begin{table*}
\centering
\caption{clusters ordered by amount of information requested}
\begin{tabular}{|c|c|p{10cm}|c|}\hline
Number&\%&Requested Data&Pseudonym\\ \hline\hline
148&57\%&Email only, or nothing&``User Management''\\ \hline
51&20\%&Mainly email and ability to publish on visitor's stream, and sometimes other details.&''Promotion''\\ \hline
61&23\%&More profile information, such as email, birthdate, location, plus sometimes the ability to publish on the visitor's stream, and/or other details.&''Profile Management''\\ \hline
\hline\end{tabular}
\label{AMOUNTOFINFOTABLE}
\end{table*}

Table \ref{FANCYROUNDTABLE} presents the distribution of site
categories among each cluster. Several interesting
observations can be made from this data. For example, just asking for
email seems to be insufficient for most social networks; only 9\% of
the ``User management'' category (i.e., asking for email only, or nothing
at all) are social networks. Also, the policy adopted by ``Promotion''
sites (requesting mainly email and the ability to publish on the
visitor's stream, and sometimes other details) is equally popular among
the four categories of JPSs (except ads which are rare in the data to
begin with). As we proceed from top to bottom in the table, 
corresponding to more information being requested, blogs become less
common, whereas social networks become more common. Shopping sites appear to be bi-modal between those
that request minimum versus maximum information (48\% v. 49\%).

\begin{table}
\centering
\caption{Distribution of categories by cluster (\%)}
\begin{tabular}{|c|c|c|c|c|} \hline
Cluster&Blog&Shop&Social&Ads\\ \hline\hline
User management&42&48&9&1\\ \hline
Promotion&29&24&24&2\\ \hline
Profile management&20&49&30&2\\ \hline
\hline\end{tabular}
\label{FANCYROUNDTABLE}
\end{table}

\begin{figure}[ht!]
\centering
\includegraphics[width=90mm]{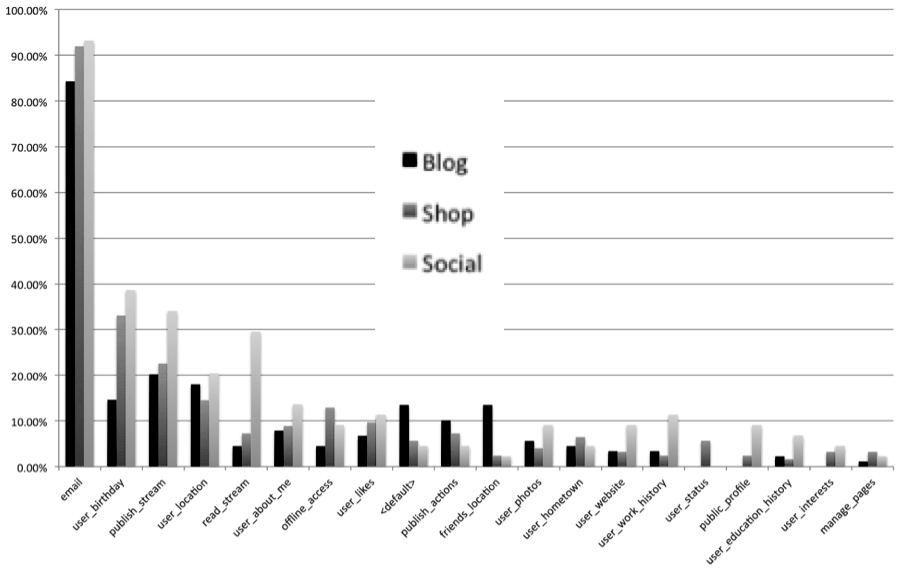}
\caption{Differential usage of information in Facebook login requests}
\label{INFORONLOGINHISTO}
\end{figure}

We also asked the opposite question: What permissions are popular in
the different web site categories? Figure \ref{INFORONLOGINHISTO}
represents the top 20 (of 48) differences (below this threshold the
difference are too small to interpret).

Social sites usually request more top Facebook information than other
categories, and defaults are not often requested by this type of
site. Defaults and publish actions are commonly requested by
blogs. Somewhat surprisingly, so is friend location. Offline access
(which is deprecated), and visitor status, are often requested by
shopping sites. Shopping sites actually vary the most for requested
information. This may be a result of natural variability in the type
of products offered.

It seemed odd to us that so many blogs were interested in friend
locations, so we examined some of them manually, and noticed that many
were political, for example: \url{don-brown-4tx14.org} (Don Brown for
congress), \url{clark-republicans.org}, \url{citizen-actionwi.org},
\url{raise-your-vote.com}, \url{save-pacifica.com}, and
\url{yes-on-1tn.org}. It makes some sense to ask for friends in order
to build a political network. A finer-grained categorization might
have put these sorts of sites into a separate category.

\pagebreak
\section{Passing Passwords in the Clear}

Actual registration is at the limit of complexity that we were able to
automatically explore. While there are many ``safe'' ways to deal with
passwords, such as sending re-pass-wording links, or sending one-time
temporary passwords, many JPSs create a password for a new user, and
then email it, unencrypted, to the user. Surprisingly, this is even
the case for some sites that use social logins. Some popular
platforms, such as Magento or Wordpress, provide tools for safe
password exchange with email templates, but their defaults are unsafe,
and JPS operators generally will not change these. We examined the
prevalence of bad password practices by inspecting emails received
after site registration: approximately 10\% of sites using custom
registration implementations sent a plain text password via email,
whereas only 2\% of those employed an activation link. Among sites
using a login button, 4\% sent a plain text password in email, versus
1\% sending an activation link. Our statistics here are somewhat
biased because we were unable to automatically answer a CAPTCHA test
or enter a separate email during the process of
registration. Regardless, these results provide a lower bound on the
surprising number of sites that employ unsafe password practices.

\section{Discussion}

\subsection{Result Highlights}

We have observed that JPS share a great deal of information with third
parties, mostly without their visitors' knowledge, and probably
without the JPS operators understanding what is being shared, nor the
implications of this sharing. When you visit a site of the sort that
we have studied, an average of 7 third party organizations find out
about you, at least in terms of your browser profile. 82\% of JPSs
send at least one request to third party sites when they load. More
than 67\% of JPSs use at least one service of Google, and 19\% use at
least one service of Facebook. Setting aside static resource usage,
which is generally considered less of a privacy concern, the most
popular third party service is analytics, used by 53\% of
JPSs. Whenever social icons appear on a page, for example a Facebook
``like'' button, Facebook may be finding out that you have visited
that page, even just on page load, without any button having to be
clicked!

JPSs are collecting a great deal of information from their community.
49\% of JPSs explicitly ask for visitors' email. When Facebook login is
used, 23\% of JPSs request full profile management permissions from
Facebook, and some are even requesting friend location information. 

JPSs permit a great deal of tracking of their community. Third party
cookies are used by more than 50\% of JPSs, and the most popular third
parties, Google and Twitter, use cookies for tracking. Both ``own''
and third party cookies usually live more than one year. Furthermore,
around 30\% of JPSs that use Facebook login ask for user\_birthday,
even given Facebook's guidance against this.

Deprecated permissions are still commonly requested, and JPSs often
fail to update to the latest privacy practices. At least 5-10\% of web
sites with Facebook login still store and send passwords
explicitly. This situation is worse with self-implemented user
management; the default Magento email template sends the password
explicitly, and the same is true for some Wordpress plug-ins.

\subsection{Limitations and Improvements}

Automatic analysis of complex web sites is very difficult, largely
because of the use of iframes, rest widgets, and
java\-script. Fortunately, our focus on JPSs ameliorates this problem
slightly, as a smaller fraction of JPSs use sophisticated methods, or
the widgetry they do use is well understood, such as third party login
methods used in the default manner recommended by the
provider. Ajax-based crawling, although much more complex, would
enable a greater range of data gathering. It would also be useful to
complement our observations with ``ground truth'', for example by way
of surveys of the operators of JPSs.

\subsection{Conclusions}

Generally speaking our analyzes support our expectation that many of
the practices of Just Plain Sites are potentially dangerous to visitor
privacy. Our goal is not to scold JPS operators, but to raise
awareness, both among JPS operators and among visitors to such
sites. Both of these constituencies would probably be surprised, if
not shocked, at what they may be inadvertently putting at risk. For
JPSs on the web, collecting private data may sometimes be important to
the site operators, and sometimes they may be aware of it -- we are
\textit{not} claiming that visiting these sites always constitutes an
\textit{actual} danger of invasion of privacy; it may well be that
such information leakage offers a benefit to the site operators and
visitors, for example, by improving targeted advertising. But much of
the time it is probably \textit{not} important, and the site operators
are \textit{not} aware of it. Compare this with top tier sites where,
even if the visitors are not aware or concerned with what is being
taken from them, it is nearly certain that the site operators are well
aware of these details. Indeed, in many cases this is their explicit
business model, and when it is not their business model to sell out
their visitors, such top tier sites have the resources to ensure that
they understand and properly implement best privacy practices. This is
not so for JPSs.  Hopefully our work can move JPS visitors a small way
towards having a richer understanding of what is going on when they
visit these sites, and at the same time move JPS operators a small way
towards awareness of potential problems, and towards properly
implementing best privacy practices.

\section{Acknowledgments}

This work was supported by CommerceNet. Several anonymous reviewers
helped us improve the presentation.

\bibliographystyle{ieeetr} \bibliography{PPoJPSs} 

\setcounter{secnumdepth}{0}
\section{Appendix: Just Plain (Mobile) Apps}

JPSs are often accessed via smartphone or tablet apps. Indeed, in some
cases this is the primary means of access. Therefore, it would make
sense to analyze mobile clients. Unfortunately, the app world is so
different from the web that we were unable to make our methods apply
without basically starting over. First off, scraping apps, as compared
to scraping web sites, is complicated, even in the case of Android's
xml-based layout of screens. Because of greater interface design
freedom, it is hard to properly detect mobile analogues of forms, and
the correspondences among positionally separated labels and
inputs. Moreover many app developers us dynamic text. On the other
hand, app user management, third party logins, analytics, and other
practices are similar to web sites. 

We tried to answer the questions of how many JPS web sites have
separate mobile versions, and how many JPSs have supporting mobile
apps. However, only about 10\% of the JPS sites that we accessed had
mobile redirection (e.g., ``/m.'', ``.mobi''). Many sites now use
``universal layout'', with separate divs, or an entirely separate site
for mobile access. We are unable to see such solutions in our existing
dataset. In a second analysis we searched Google Play for apps
matching each JPS web site by name. The percentage was very low as
well, perhaps because in many cases, as mentioned, the developers of
the app for a JPS are not the same as the JPS. These preliminary
analyses yielded low rates and difficult-to-interpret results.

Fortunately for the user, mobile developers have to try somewhat
harder to get data off the phone or pad than than when a small
mom-and-pop shop puts up their own web site, so mobile developers are
somewhat more likely to know what they are doing in this regard, or at
least know that they are doing it. Also, apps that are distributed by
online stores, such as iTunes and Google Play, are subject to fairly
strict policies regarding data collection and privacy.

For all these reasons we put our analysis of JPS apps (JPAs?) on hold
for another time.

\end{document}